\documentclass[aps,prl,preprint,superscriptaddress]{revtex4-1}

\usepackage[version=3]{mhchem} 
\usepackage[T1]{fontenc}       
\usepackage{graphicx}
\usepackage{subfigure}
\usepackage{natbib}
\usepackage{url}
\usepackage{amsmath}
\usepackage{amssymb}
\usepackage{mathrsfs}
\usepackage{color}
\usepackage{ulem} 
\usepackage{soul,xcolor}
\usepackage{comment}
\usepackage[english]{babel}
\emergencystretch=\maxdimen
\providecommand\Oh{\mathit{Oh}}

\begin{document}


\title{Self-Propelled Detachment upon Coalescence of Surface Bubbles}



\author{Pengyu Lv}
\affiliation{State Key Laboratory for Turbulence and Complex Systems, Department of Mechanics and Engineering Science, BIC-ESAT, College of Engineering, Peking University, Beijing 100871, People's Republic of China}
\affiliation{Physics of Fluids group, Faculty of Science and Technology, Max Planck - University of Twente Center for Complex Fluid Dynamics, MESA+ Institute, and J. M. Burgers Centre for Fluid Dynamics, University of Twente, P.O. Box 217, 7500 AE Enschede, Netherlands}

\author{Pablo Pe\~{n}as}
\affiliation{Physics of Fluids group, Faculty of Science and Technology, Max Planck - University of Twente Center for Complex Fluid Dynamics, MESA+ Institute, and J. M. Burgers Centre for Fluid Dynamics, University of Twente, P.O. Box 217, 7500 AE Enschede, Netherlands}

\author{Hai Le The}
\affiliation{Physics of Fluids group, Faculty of Science and Technology, Max Planck - University of Twente Center for Complex Fluid Dynamics, MESA+ Institute, and J. M. Burgers Centre for Fluid Dynamics, University of Twente, P.O. Box 217, 7500 AE Enschede, Netherlands}
\affiliation{BIOS Lab-on-a-Chip group, Faculty of Electrical Engineering, Max Planck - University of Twente Center for Complex Fluid Dynamics, Mathematics and Computer Science, MESA+ Institute, University of Twente, P.O. Box 217, 7500 AE Enschede, Netherlands}

\author{Jan Eijkel}
\affiliation{BIOS Lab-on-a-Chip group, Faculty of Electrical Engineering, Max Planck - University of Twente Center for Complex Fluid Dynamics, Mathematics and Computer Science, MESA+ Institute, University of Twente, P.O. Box 217, 7500 AE Enschede, Netherlands}

\author{Albert van den Berg}
\affiliation{BIOS Lab-on-a-Chip group, Faculty of Electrical Engineering, Max Planck - University of Twente Center for Complex Fluid Dynamics, Mathematics and Computer Science, MESA+ Institute, University of Twente, P.O. Box 217, 7500 AE Enschede, Netherlands}

\author{Xuehua Zhang}
\email{xuehua.zhang@ualberta.ca}
\affiliation{Department of Chemical \& Materials Engineering, University of Alberta,
Edmonton, Alberta, T6G1H9, Canada}
\affiliation{Physics of Fluids group, Faculty of Science and Technology, Max Planck - University of Twente Center for Complex Fluid Dynamics, MESA+ Institute, and J. M. Burgers Centre for Fluid Dynamics, University of Twente, P.O. Box 217, 7500 AE Enschede, Netherlands}

\author{Detlef Lohse}
\email{d.lohse@utwente.nl}
\affiliation{Physics of Fluids group, Faculty of Science and Technology, Max Planck - University of Twente Center for Complex Fluid Dynamics, MESA+ Institute, and J. M. Burgers Centre for Fluid Dynamics, University of Twente, P.O. Box 217, 7500 AE Enschede, Netherlands}
\affiliation{Max Planck Institute for Dynamics and Self-Organization, 37077 G\"ottingen, Germany}


\date{\today}

\begin{abstract}
The removal of microbubbles from substrates is crucial for the efficiency of many catalytic and electrochemical gas evolution reactions in liquids. The current work investigates the coalescence and detachment of bubbles generated from catalytic decomposition of hydrogen peroxide. Self-propelled detachment, induced by the coalescence of two bubbles, is observed at sizes much smaller than those determined by buoyancy.  Upon coalescence, the released surface energy is partly dissipated by the bubble oscillations,  working against viscous drag.  The remaining energy  is converted to the kinetic energy of the out-of-plane jumping motion of the merged bubble. The critical ratio of the parent bubble sizes for the jumping to occur is theoretically derived from an energy balance argument and found to be in agreement with the experimental results.
The present results provide both physical insight for the bubble interactions and practical strategies for applications in chemical engineering and renewable energy technologies like electrolysis.

\end{abstract}


\maketitle


Gas-evolution reactions are part of many electrochemical processes employed in industry, and importantly hold the promise of delivering sustainable energy technologies such as solar hydrogen \cite{sorrell2002,ardo2018}.
Tremendous research efforts have focused on understanding and fabrication of efficient photocatalysts for gas evolution reactions \cite{jakob2012,lu2014ultrahigh,jiang2015},  yet
electrochemical systems often still suffer from poor bubble management \cite{Taqieddin2017}.
Bubbles on the catalyst surface reduce the effective contact area between the catalyst and the reacting liquid, thereby increasing the  electric resistances \cite{takanabe2017photocatalytic, shen2008photocatalytic}. In contrast, promoting bubble detachment can strengthen the local liquid convection and renewal \cite{lakkaraju2013heat,guzman2016,guzman2016b,sur2018pool}, consequently increasing the energy efficiency of the reaction  \cite{wilhelmsen2015heat, cho2015turning, whitney1988mass}. It is then evident that rational design of gas-evolving electrochemical processes requires good control over the growth and transport of the evolved bubbles \cite{lohse2018}.

The adhesion of a bubble on a solid surface is provided by the capillary force exerted at the bubble rim (determined by interfacial tensions and wettability of the electrode surface) and countered by buoyancy. The so-called Fritz radius, i.e., the maximum detachment radius of a spherical bubble, quantifies adhesion. It is  theoretically obtained from a
buoyancy and capillary force balance \cite{oguz1993dynamics, duhar2006dynamics}:
\begin{equation}
R_F = \left( \frac{3R_p \sigma}{2 g \Delta\rho }\right)^{1/3},
\label{eq1}
\end{equation}
where $ \Delta \rho$ is the density difference between the gas and liquid, $\sigma$ surface tension, $g$ gravity, and $R_p$ the contact radius. Ultrahigh performance of electrolysis has been shown possible upon minimisation of the capillary force, e.g., on  electrodes with underwater super-aerophobicity (large wettability, small $R_p$) which accelerate bubble detachment \cite{lu2014ultrahigh}. Another method is preferential bubble formation by providing hydrophobic domains, e.g., located outside of the pore space in a porous electrode \cite{kadyk2016}.

Bubble coalescence dramatically changes the stability of individual bubbles  \cite{li2014coalescence, chan2015coalescence, zhang2006removal, shin2015growth, lv2017morphological, xue2015morphology, thoroddsen2005, zhang2008satellite, soto2018coalescence} -- to the extent that  Eq.\ (\ref{eq1}) may become impractical -- and therefore plays a key role in the evolution of the bubble population on surfaces. An analog process is droplet coalescence on surfaces. The coalescence of similarly-sized droplets may lead to out-of-plane jumping motion of the merged droplet \cite{boreyko2009self}, which is a consequence of  the release of excess surface energy upon coalescence atop nanostructured surfaces \cite{boreyko2009self, zhang2009satellite, wisdom2013self, liu2014numerical, enright2014coalescing, mouterde2017antifogging, mouterde2017merging, lecointre2019ballistics}.


Previous work has shown that when two bubbles of different sizes coalesce, the center of the merged bubble shifts on the plane towards the larger parent bubble \cite{weon2012coalescence, kim2015coalescence, weon2008coherent, chen2017spatial}, or that the merged bubble stays pinned to the surface provided its radius $R_m < R_F$ \cite{soto2018coalescence}. However, it will be shown in this paper that coalescence can also result in a "jumping mode", namely in the self-propelled detachment of merged bubbles, whose premature sizes lie well below $R_F$. The ability to predict and promote jumping bubbles is thus of extreme practical relevance.
To this end, we investigate bubble relocation and detachment triggered by the coalescence of two neighboring bubbles. In particular, we identify the threshold in parameter space beyond which a merged bubble will prematurely jump off the surface.

\begin{figure}[ht]
\centering
\includegraphics[width=12cm]{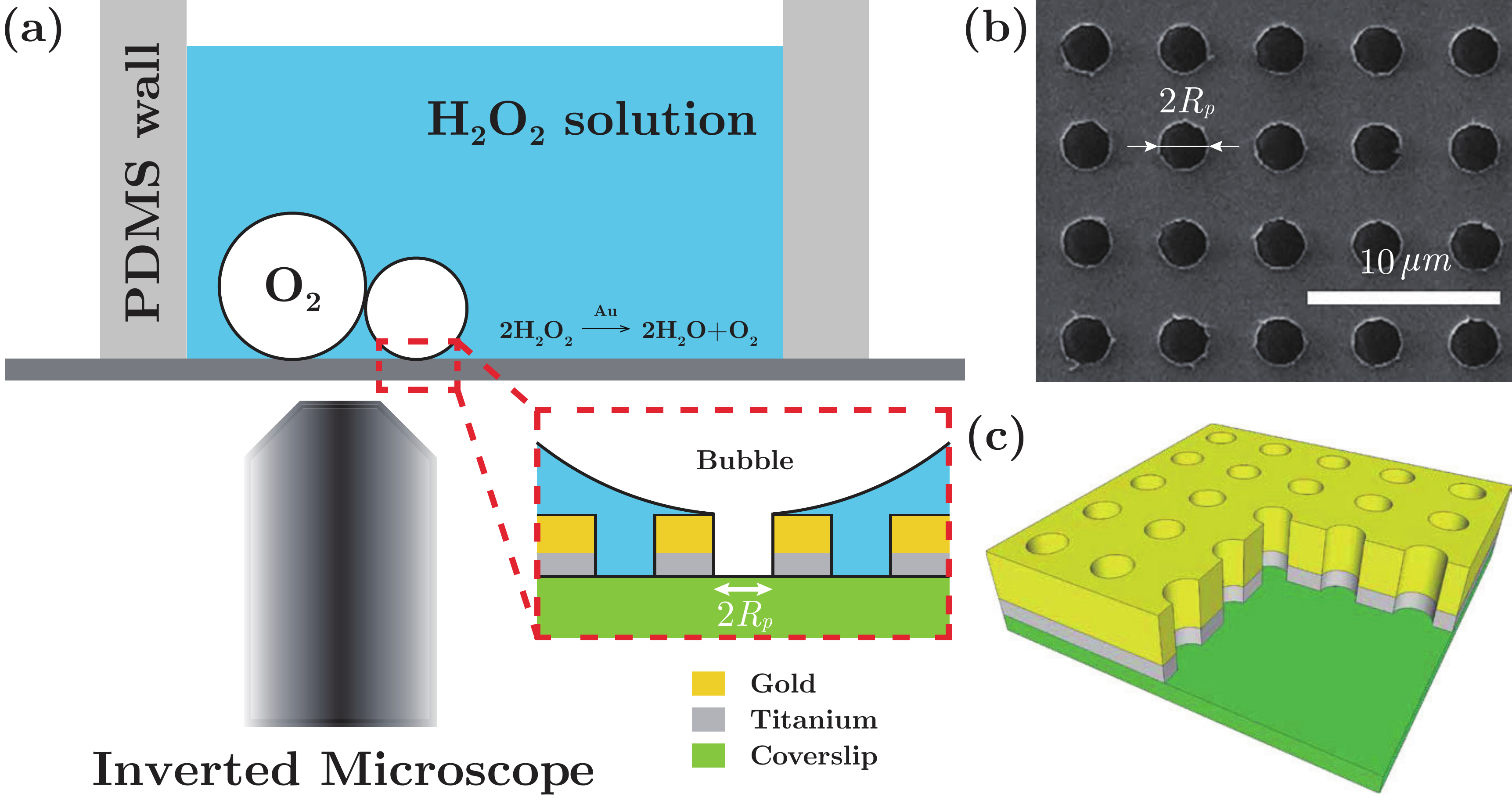}
\caption{(a) Schematics of the experimental setup. A PDMS container sticking to a glass coverslip is filled with 30\% (w/w) \ce{H2O2} solution with a volume of $10 \times 10 \times 7.5$ $\mathrm{mm^3}$. The coverslip with a thickness of 170 $\mu$m is coated with a 18-nm thick gold-titanium binary layer (15 nm for gold layer and 3 nm for titanium layer) patterned with micro-holes. The decomposition of \ce{H2O2} is taking place at the gold-solution interface, followed by the formation of bubbles which pin at the edge of the micro-holes (see inset). The bubble dynamics is measured by a high-speed camera and confocal microscopy. (b) Scanning electron microscopic image showing the hole-patterned binary layer. Each hole has a radius of $R_p = 1$ $\mu$m and the center distance is 5 $\mu$m. The scale bar is 10 $\mu$m. (c) Three dimensional schematics of the hole-patterned catalyst substrate.}
\label{Experimental}
\end{figure}

Our experiments concern oxygen bubbles that evolve from the catalytic decomposition of hydrogen peroxide (\ce{H2O2}) on a catalyst surface. The experimental setup is shown in Fig.~\ref{Experimental}(a). Gold was selected to be the catalyst and exposed to \ce{H2O2} solution to accelerate the generation of oxygen. A micro-hole patterned gold-titanium binary layer [Fig.~\ref{Experimental}(b) and~\ref{Experimental}(c)] was fabricated on a glass substrate (170 $\mu$m thick) using conventional photolithography and lift-off processes. This hole pattern provides plenty of orderly distributed sites for bubble nucleation, which enables the controlled formation and coalescence of bubbles. The gold layer ($\sim15$ nm thick) and a titanium layer ($\sim3$ nm thick) were sputtered by using an ion-beam sputtering system (home-built T'COathy system, MESA+, NanoLab) \cite{le-the2018large-scale}, where titanium was used to strengthen the binding performance between gold and glass in order to avoid any delamination during bubble formation. The static, advancing, and receding contact angles on the micro-hole patterned gold surface were
 $85.3\pm4.0^{\circ}$, $89.7\pm2.0^{\circ}$, and $19.6\pm3.6^{\circ}$, respectively.
The bubbles nucleated and grew out of an oxygen-oversaturated environment \cite{lv2017growth, chen2009growth}, being pinned at the edge of the micro-holes with a footprint radius $R_p = 1$ $\mu$m, which was measured by an inverted laser scanning confocal microscope (A1 system, Nikon Corporation, Japan). The bubble growth and interactions were recorded using a high-speed camera (Phantom V2512, USA) through an inverted microscope with a dry objective (Olympus, UPLFLN 20$\times$, NA = 0.5, WD = 2.1 mm). The microscope was always focused on the gold-solution interface from the bottom view.
The bubble projections were captured at 100 fps, sufficient to determine the bubble locations and sizes immediately before and after coalescence.


\begin{figure}[ht]
\centering
\includegraphics[width=12cm]{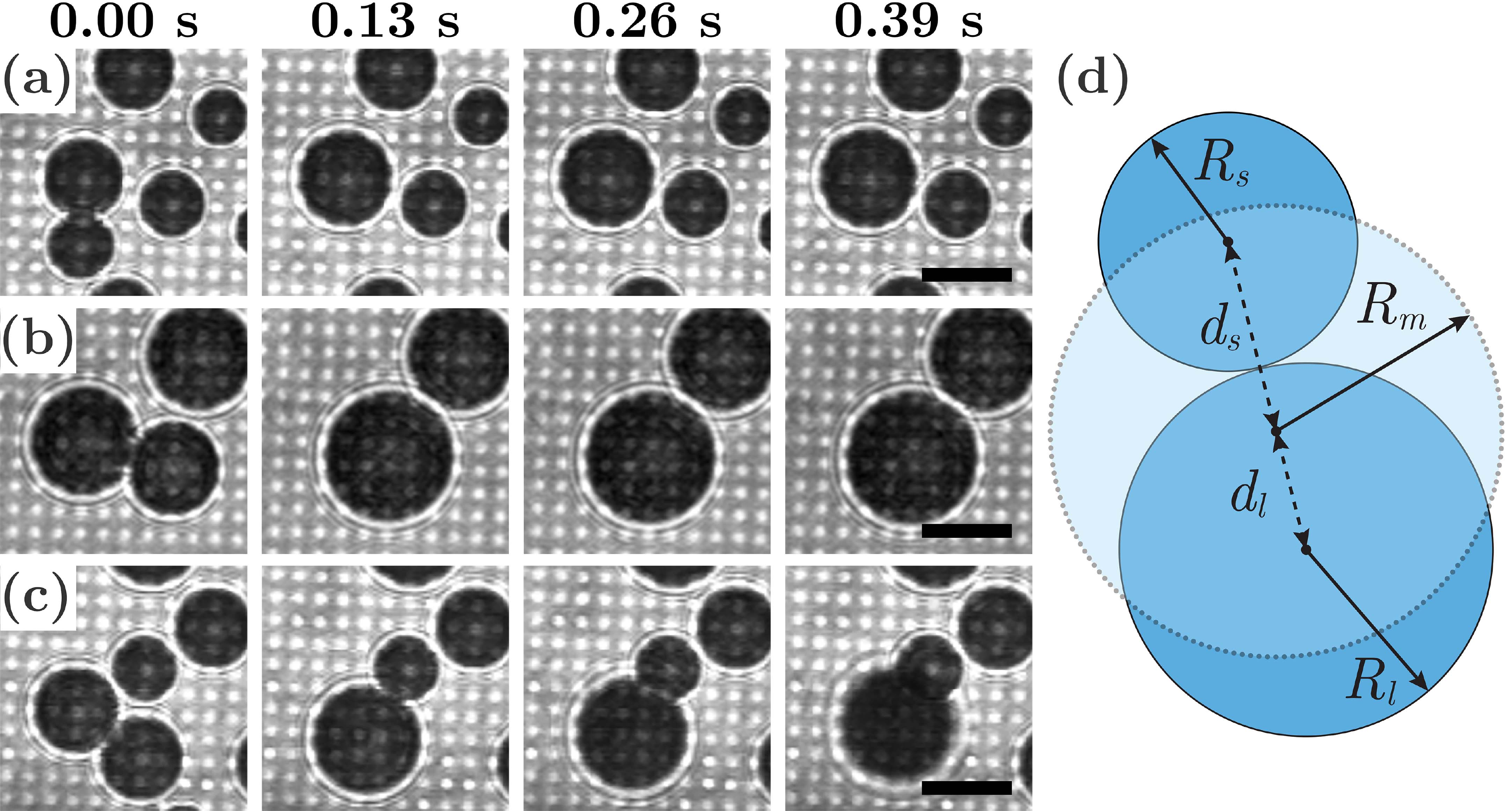}
\caption{Snapshots of the coalescence event between two parent bubbles resulting in a (a) sticking, (b) moving, or (c) jumping merged bubble. Coalescence occurs at 0.13 s.   The sticking merged bubble in (a) sticks to the location of the large parent bubble; the moving and jumping bubbles in (b, c) relocate to the center of mass of the parent bubbles.
Note that the merged bubble in (c) is out of focus after coalescence, indicating jumping followed by rising due to buoyancy. The scale bars in (a), (b) and (c) are all 20 $\mu$m. See supplemental video. (d) Schematic of the coalescence geometry and notation. The parent bubbles are shaded in dark blue; the resulting merged bubble in light blue.}
\label{BubbleCoalescence}
\end{figure}

\begin{figure}[ht]
\centering
\includegraphics[width=12cm]{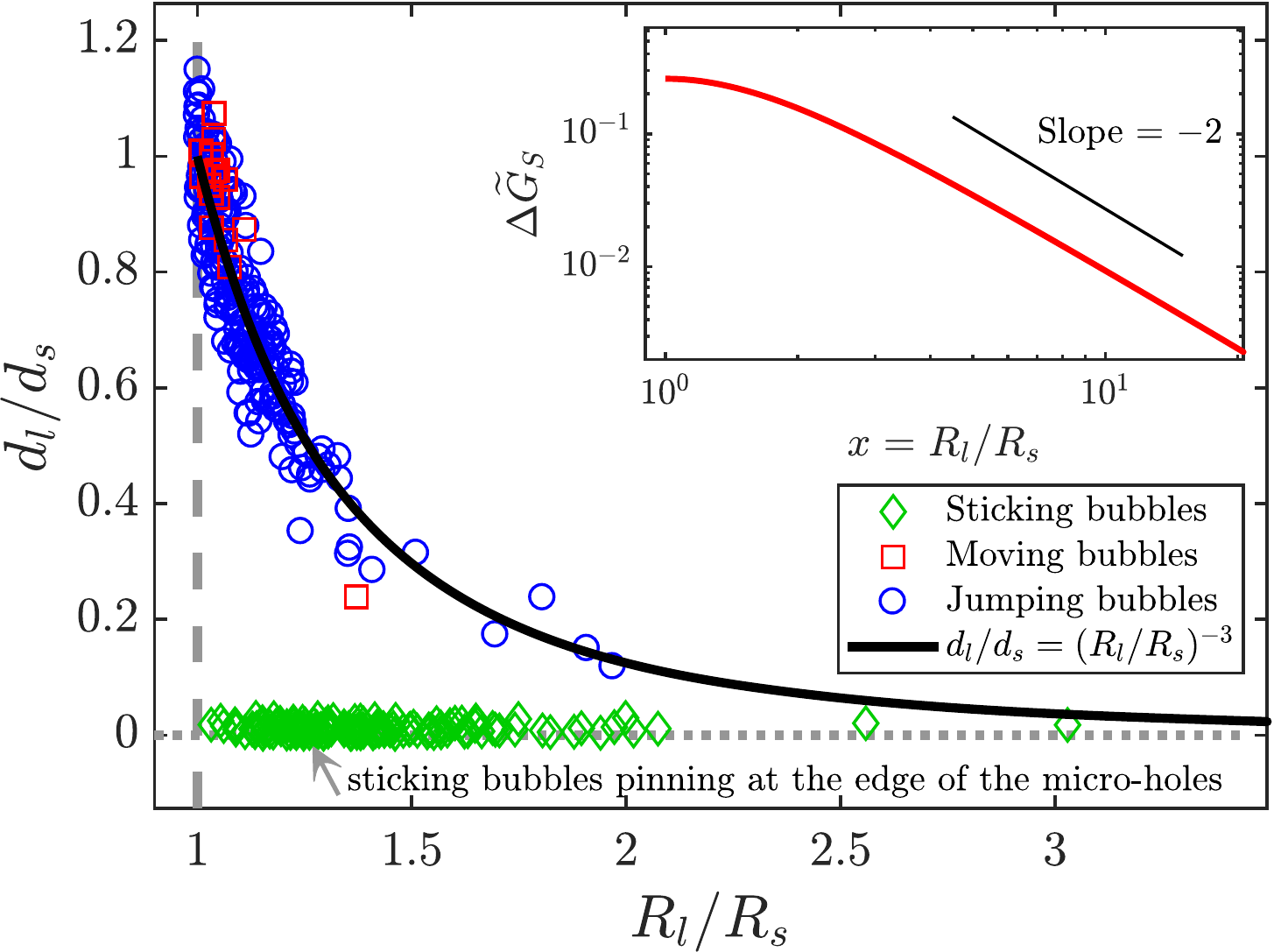}
\caption{Coalescence preference showing the relative position of the merged bubble as a function of the parent size ratio. Experimental data (symbols) were collected immediately after coalescence. The sticking bubbles pin at the location of the larger parent bubble, resulting in the data with $d_l / d_s \approx 0$.
The solid line shows the theoretical prediction Eq.\ (\ref{ScalingCoalescencePreference}) following from momentum conservation. The inset shows a double logarithmic plot of the released surface energy for the jumping bubbles as a function of the parent size ratio $x = R_l / R_s$, see Eq.~(\ref{SurfaceEnergyRelease}). The released surface energy increases as the size ratio decreases, which favors the jumping motion of the merged bubble.
 }
\label{CoalescencePreference}
\end{figure}

Figure~\ref{BubbleCoalescence} shows the evolution and coalescence of two neighboring bubbles in bottom view. Immediately after coalescence, three typical behaviors of the merged bubble were captured: (i) Sticking to the very same position on the substrate as the larger parent bubble, termed "sticking bubble" [Fig.~\ref{BubbleCoalescence}(a)]; (ii) Moving to the center of mass of the parent bubbles but still adhering to the substrate: "moving bubble" [Fig.~\ref{BubbleCoalescence}(b)]; (iii) Moving to the center of mass of the parent bubbles and spontaneously jumping off: "jumping bubble" [Fig.~\ref{BubbleCoalescence}(c)]. To describe the relative position of the merged bubble immediately after coalescence, $d_l$ and $d_s$ are defined as the distances from the merged center to the center of the larger (radius $R_l$) and the smaller (radius $R_s$) parent bubbles, respectively [see Fig.~\ref{BubbleCoalescence}(d)]. The relationship between the relative position of the merged bubble $d_l / d_s$ and the parent size ratio $R_l / R_s$ reveals the coalescence preference \cite{weon2012coalescence}, see Fig.~\ref{CoalescencePreference}, in which our experimental data are shown. On the one hand, for sticking bubbles, especially the ones with larger parent size ratios, the center will indeed stick to the location of the larger parent, which is reflected by $d_l / d_s \approx 0$.
Here, the contact line pinning force acting on the edge of the micro-hole of the larger parent bubble is large enough to control momentum conservation during coalescence.
On the other hand, for both moving and jumping bubbles, $d_l \neq 0$ and the relative position $d_l / d_s$ decreases as the parent size ratio $R_l / R_s$ increases, indicating that pinning loses relevance and that the merged bubble is preferentially closer to the larger parent. This coalescence preference can be in fact attributed to the conservation of momentum in the horizontal plane, overcoming the pinning forces. Upon coalescence, the parent bubbles will move towards each other while the horizontal momentum of the system must be conserved throughout coalescence, leading to $m_l u_l(t) - m_s u_s(t) = 0$, where $m_l$ ($m_s$) and $u_l(t)$ ($u_s(t)$) are the added mass \cite{magnaudet2000} and the horizontal velocity of the larger (smaller) parent bubble, respectively. The added masses of the parent bubbles can be considered unchanged during coalescence, as the whole coalescence process is much faster than the diffusive growth of the bubbles.
Noting that $d_l = \int_{t_1}^{t_2} u_l(t) \mathrm{d}t$ and $d_s = \int_{t_1}^{t_2} u_s(t) \mathrm{d}t$, where $t_1$ and $t_2$ are the moments immediately before and after coalescence, respectively, one may integrate the horizontal momentum balance in time to find that $m_l d_l = m_s d_s$. In other words, the merged bubble associated with the moving and jumping modes is horizontally located at the projection of the center of mass of the parent-bubble system. Therefore, the relative position of the merged bubble,
\begin{equation}
\frac{d_l}{d_s} = \left( \frac{m_l}{m_s} \right) ^ {-1} = \left( \frac{R_l}{R_s} \right) ^ {-3},
\label{ScalingCoalescencePreference}
\end{equation}
is found to depend only on the parent size ratio. We plot the theoretical result Eq.\ (\ref{ScalingCoalescencePreference}) in Fig.~\ref{CoalescencePreference} together with the experimental data, finding excellent agreement, without any fitting parameter. This agreement  demonstrates that the coalescence preference simply follows from the conservation of horizontal momentum, reflecting that pinning forces at the edge of the hole are indeed irrelevant. However, for the moving bubbles, the pinning site after coalescence usually offsets from the projection of the center of mass, consequently leading to some scatter of  $d_l/d_s$. In addition, as plenty of bubbles are formed in a cluster, disturbances from the bubble growth, coalescence, and detachment in the surroundings will affect the movement of the merged bubbles, also resulting in scatter of the data for moving bubbles and jumping bubbles.

In contrast to our results, Weon et al. \cite{weon2012coalescence} reported a scaling exponent of $-5$ rather than $-3$ as we find in Eq.\ (\ref{ScalingCoalescencePreference}), which was attributed to the topological change caused by the release of excess surface energy. It is striking that in our experiments surface energy seems to have  no evident effect on the coalescence preference or position, as reflected in
 the agreement with the center-of-mass theory. Nonetheless, the discrepancy between the --3 and --5 scaling
 may be explained by comparing both coalescence processes. Firstly, our moving bubbles are subject to a pinning effect: the merged bubble is drawn to the nearest micro-hole after coalescence. The remaining surface energy release is expectedly dissipated in the flow of the surrounding viscous liquid. In Ref. \cite{weon2012coalescence}, there is no pinning effect,
  given that the bubbles can move freely on the oil-water interface. Secondly, our jumping bubbles present out-of-plane motion.  In contrast, in Ref.\  \cite{weon2012coalescence} the mobility of the merged bubbles was  constrained to in-plane movement by the interface and therefore no jumping was observed.
Thus in our configuration, the excess surface energy release is  eventually converted to the kinetic energy of the bubble jump (i.e., to out-of-plane motion),  whereas in Ref. \cite{weon2012coalescence}, it only contributes to the in-plane momentum and in-plane motion.

In our experiments, the excess surface energy is so large that the maximum radius of the merged jumping bubbles
 is 64 $\mu$m, i.e., less than 30\% of the calculated Fritz radius $R_F = 217$ $\mu$m (taking $R_p = 1$ $\mu$m) and less than 3\% of the corresponding volumes. Coalescence, not buoyancy, is therefore responsible for the jumping motion.

Coalescence generates capillary waves that propagate along the bubble surface and meet at the opposite apex of the coalescence point \cite{soto2018coalescence}; such convergence takes place over the capillary time scale $T_c = \sqrt{\rho R_m^3/\sigma}$, where $\rho$ is the liquid density and $R_m$ the radius of the merged bubble \cite{lamb1945Hydrodynamics,zhang2008satellite,soto2018coalescence}. Surface tension then drives the retraction of low-mode capillary waves (visibly corresponding to shape oscillations of mode two \cite{soto2018coalescence}) towards a spherical shape. Jumping occurs when the retraction process is vigorous enough. Note that higher modes of oscillation (shorter wavelengths) are fully damped by viscosity during the retraction, as explained by the fact that the viscous dissipation time is proportional to the square of the wavelength $\lambda$ \cite{lamb1945Hydrodynamics}, $T_\mu = \rho \lambda^2/\mu$, with $\mu$ the dynamic viscosity.



The condition for jumping can be semi-quantitatively estimated by means of scaling arguments and a simple energy balance. The core argument is that, for permanent detachment or jumping to occur, the excess surface energy must overcome the work done by the bubble against the viscous drag during the expansion and retraction process. We begin by assuming that the bubble volume is conserved before and after coalescence. The bubble shape is considered to be a perfect sphere, as the pinning area ($R_p = 1$ $\mu$m) is negligible. Volume conservation implies that
$R_m = (1 + x^3)^{\frac{1}{3}} R_s$, with the parent size ratio $x = R_l / R_s$. The released surface energy after coalescence is
\begin{equation}
\Delta G_S = 4 \pi \sigma \left( R^2_{l} + R^2_{s} - R^2_{m} \right).
\label{SurfaceEnergy}
\end{equation}
The adhesion energy, which for all bubbles only depends on the radius of the pinning site,
 can be neglected given that it is small compared to surface energy released:
\begin{equation}
\frac{W_a}{\Delta G_s} \approx  \frac{\pi \sigma R_p^2 }{4\pi \sigma R_m^2 \Delta \widetilde G_s} \leq O(0.1),
\end{equation}
where we have taken $R_p/R_m \sim 0.1$ and $\Delta \widetilde G_s \sim 0.1$ for $x$ of order unity (cf. inset of Fig.~\ref{CoalescencePreference}).

During the low-mode shape oscillations after coalescence, the functional dependence of the viscous drag exerted on the merged bubble is assumed to be Stokes-like, as expected for a bubble undergoing pure translational or volumetric oscillations \cite{marmottant2006high}:
\begin{equation}
F_\mu \sim \mu R_m U.
\label{ViscousDrag}
\end{equation}
The velocity scale ${U} \sim \varepsilon \omega_m R_m$ represents the characteristic velocity of the retraction process (low-mode oscillation) with $\omega_m \sim \sqrt{\sigma/\rho R_m^3} = 1/T_c$  being the natural frequency of low-mode shape oscillations \cite{versluis2010microbubble} and $\varepsilon$ a dimensionless amplitude which takes into account viscous damping. We expect $\varepsilon$ to be a decreasing function of $\mu$, meaning that a higher viscosity implies a lower retraction speed.  And $\varepsilon$ must also depend on the initial deformation (excess surface energy). Taking the retraction distance to be proportional to $R_m$, the retraction distance has to scale with the excess length due to coalescence (amplitude of deformation), and not with the merger bubble size. Then the work
$W_\mu$ done against viscous drag consequently scales as
\begin{equation}
W_\mu \sim F_\mu R_m \sim  \varepsilon\mu \sqrt{\sigma/\rho} R_m^\frac{3}{2}.
\label{Work}
\end{equation}
Energy conservation during coalescence results in  $\Delta G_S = W_\mu + E_k$, with $E_k$ the kinetic energy available for the  jump. This implies
 that the bubble will not jump unless $\Delta G_S > W_\mu$; thus, the critical condition is $\Delta G_S = W_\mu$. Equating (\ref{SurfaceEnergy}) and (\ref{Work}) allows to express the critical radius of the larger parent bubble $R_{l,cr}$ as a function of the parent size ratio $ x = R_l / R_s$. We obtain that the smallest radius for which jumping occurs is
\begin{equation}
R_{l,cr}(x) = R^* \frac{x \left( 1 + x^3 \right)}{\left[ 1 + x^2 - \left( 1 + x^3 \right) ^ \frac{2}{3} \right] ^ 2}.
\label{CriticalRadius}
\end{equation}
Here, $R^* \propto \mu^2 \varepsilon^2/\sigma\rho$  represents a coefficient of proportionality with dimensions of length that absorbs all numerical prefactors and depends only on the liquid properties and $\varepsilon$.

\begin{figure}
\centering
\includegraphics[width=16cm]{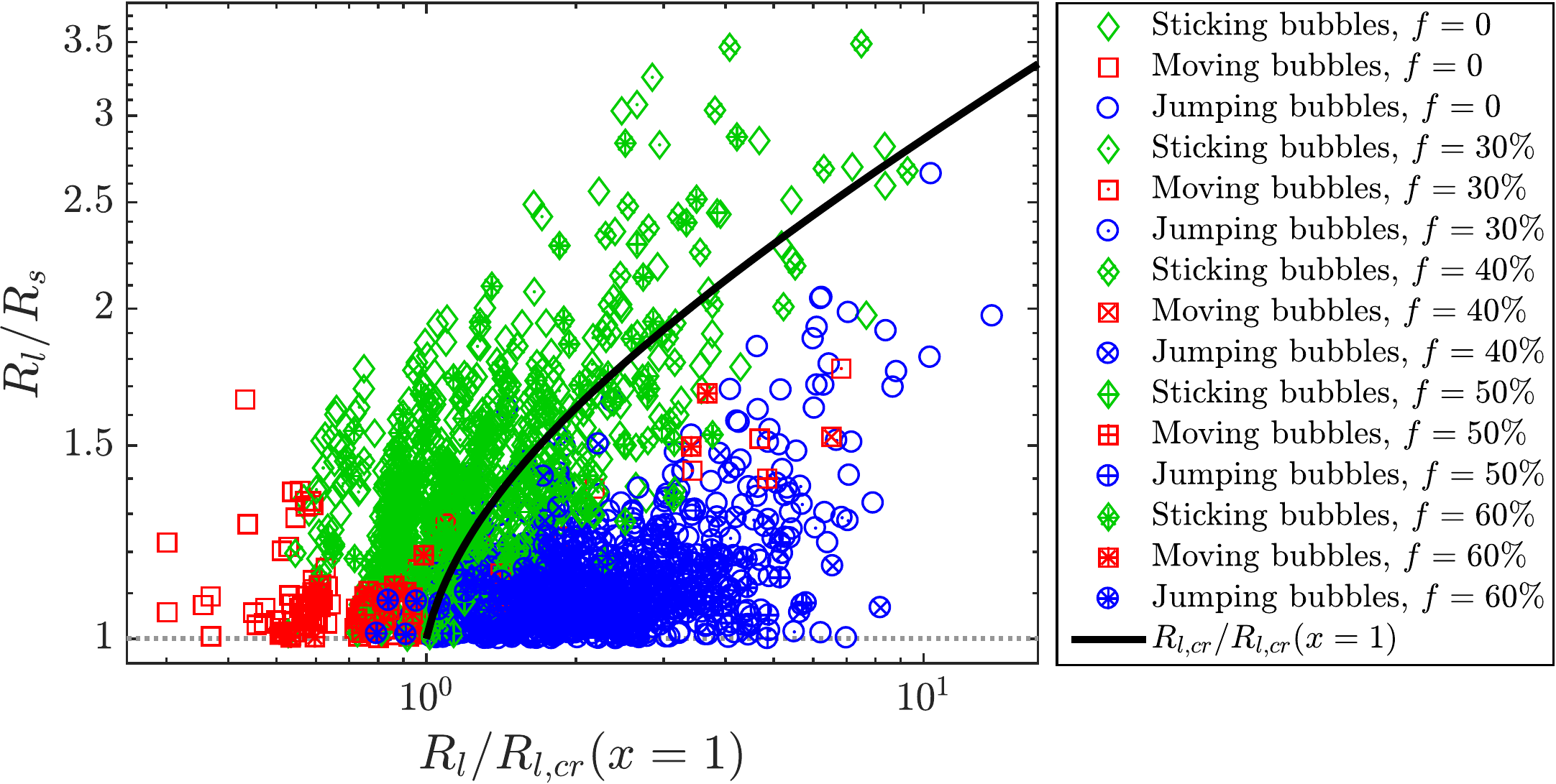}
\caption{Phase diagram of the motion regimes for the merged bubbles. The theoretical threshold curve (solid line, Eq.\ (\ref{CriticalRadius})) divides the phase diagram into two regimes, namely non-jumping (to the left), including sticking and moving, and jumping (to the right) motions upon coalescence. Experimental data (symbols) were collected immediately before coalescence. $f$ is the mass fraction of glycerol in the 30\% (w/w) \ce{H2O2} solution.}
\label{PhaseDiagram}
\end{figure}

Figure~\ref{PhaseDiagram} displays the phase diagram of the motion regimes for the merged bubbles, in which the experimental data for both non-jumping (sticking and moving) and jumping motions are shown.
The parents of the non-jumping bubbles usually have smaller sizes, but larger size ratios, than those of the jumping bubbles. For a fixed size of the larger bubble, the smaller the parent size ratio, the easier for the merged bubble to jump.
For parent bubbles of equal size, the critical larger parent radius is $R_{l,cr}(x = 1) = 11$ $\mu$m, which is
 directly read off from the experimental data. The full threshold curve based on Eq.~(\ref{CriticalRadius}), namely $R_{l,cr}(x)/ R_{l,cr} (x = 1)$, is also plotted in Fig.~\ref{PhaseDiagram}. The threshold curve, which is purely a function of $x$ (it does not depend on $R^*$), identifies the jumping and non-jumping behavior in the phase space determined by the parent size ratio $x = R_l / R_s$ and the normalised larger parent size $R_l/ R_{l,cr} (x = 1)$. To the right of the curve, the merged bubbles jump off ($\Delta G_S > W_\mu$); to the left, they stick to the surface ($\Delta G_S < W_\mu$).
Close to $\Delta G_S = W_\mu$, little energy is left for the bubble to jump. Nonetheless, as long as pinning forces are overcome, the bubble will detach from the substrate and rise due to buoyancy. The experimental data are in reasonable agreement with our simple estimate. However, for some larger bubble size ratios $R_l / R_s > 1.3$, the estimate Eq.~(\ref{CriticalRadius}) overpredicts jumping, presumably because we neglected the energy of capillary waves needed to overcome the pinning effect.

Note that the viscous dissipation of energy of the low surface modes $\lambda \sim R_m$ governing the retraction process has been neglected. The energy of a capillary wave with $\lambda \sim R_m$ will have decayed over a time $T_c$ by a factor $\sim \exp(-T_c/T_\mu)$, where $T_c/T_\mu   = \mu/\sqrt{\rho R_m \sigma} = \Oh_m$ is in fact the Ohnesorge number of the merged bubble.
In our experiments, taking $\sigma = 74$ mN/m and $\mu = 1.07$ mPa s for a 30\% (w/w) \ce{H2O2} solution \cite{phibbs1951hydrogen} together with a typical bubble size of $R_m \sim 50$ $\mu$m,  results in $ \Oh_m =O(10^{-2})$. Hence, the viscous decay of the low surface modes is indeed small.

To further investigate the effect of liquid viscosity on the detachment of coalesced bubbles, we added glycerol to the 30\% (w/w) \ce{H2O2} solution to change the liquid viscosity by varying mass fractions ($f=$ 30\%, 40\%, 50\%, 60\%) and performed the identical experiments. The experimental results presenting the corresponding phase diagrams are shown in the Supplemental Material \cite{Suppl}. Irrespective of the value of the viscosity, $R_{l,cr} (x = 1)$ always remains virtually the same as that of the case of pure \ce{H2O2} solution (11 $\mu$m). Moreover, the phase diagrams are all remarkably similar and in reasonable agreement with the threshold curve based on Eq.~(\ref{CriticalRadius}). We then plot all the data together in Fig.~\ref{PhaseDiagram}, indicating that viscosity has no discernible role on the critical radius. This suggests that the retraction velocity roughly scales as $\varepsilon \propto 1/\mu$. This is, to an extent, expected, especially if we think of the bubble as a damped harmonic oscillator with a damping constant  $\propto \mu/\rho R^2_m$ \cite{plesset1977bubble, stover1997bubble}. Consequently, the quantity $\varepsilon \mu$, hence $W_\mu$ and ultimately $R^*$ in  Eq.(\ref{CriticalRadius}), all remain remarkably insensitive to the liquid viscosity.

Finally, to better reveal the effect of the parent size ratio on the jumping behavior, we examine the dimensionless surface energy release,
\begin{equation}
\Delta \widetilde{G}_S = \frac{\Delta G_S}{4 \pi \sigma R_m^2} = \frac{1 + x^2 - \left( 1 + x^3 \right) ^ \frac{2}{3}}{\left( 1 + x^3 \right) ^ \frac{2}{3}},
\label{SurfaceEnergyRelease}
\end{equation}
which only depends on the parent size ratio $x$, see inset of Fig.~\ref{CoalescencePreference}. $\Delta \widetilde{G}_S$ significantly decreases as the parent size ratio increases, exhibiting a scaling relationship of $\Delta \widetilde{G}_S \propto x ^ {-2}$, which confirms that greater size inequality is associated with less excess surface energy for any jumping motion.
This outcome, consistent with the experimental results in Fig.~\ref{PhaseDiagram}, constitutes an important guideline for the efficient removal of surface-covering bubbles.

In conclusion, we report a self-propelled bubble jump resulting from the coalescence of neighboring bubbles. Bubble detachment occurs at sizes much smaller than those determined by buoyancy. The released surface energy upon coalescence is argued to be partly dissipated by the oscillations of the merged bubble  in the form of work against viscous drag, and any remaining energy is then supplied to the jumping motion. Finally, we have derived an expression for the critical bubble radius above which  jumping occurs. Further experiments and numerical simulations are required to further confirm
and supplement the developed picture. Such
line of research
will facilitate predictions and strategies for efficient removal of bubbles from surfaces in practical applications such as in catalysis, electrolysis, boiling, or plasmonic bubble generation.
In those cases, an even further enrichment of the observed phenomena may arise from the physicochemical hydrodynamics of the bubbles far from equilibrium \cite{lohse2020}, such as solutal and thermal Marangoni forces.

\begin{acknowledgments}
We thank two very insightful anonymous referees for excellent suggestions and for making us improve the paper considerably. This work was supported by the Netherlands Center for Multiscale Catalytic Energy Conversion (MCEC), an NWO Gravitation programme funded by the Ministry of Education, Culture and Science of the government of the Netherlands, and by the  ERC Advanced Grant ``DDD''  under the project number 740479. This research was undertaken, in part, thanks to funding from the Canada Research Chairs Program, and Future Energy Systems (Canada First Research Excellence Fund).
\end{acknowledgments}

\bibliographystyle{prsty_withtitle}
\bibliography{JumpingBubble}

\begin{thebibliography}{10}

\bibitem{sorrell2002}
T. Bak, J. Nowotny, M. Rekas, and C. Sorrell, {\em Photo-electrochemical
  hydrogen generation from water using solar energy. Materials-related
  aspects}, International Journal of Hydrogen Energy {\bf 27},  991   (2002).

\bibitem{ardo2018}
S. Ardo, D.~F. Rivas, M.~A. Modestino, V.~S. Greiving, F.~F. Abdi, E.~A. Llado,
  V. Artero, K. Ayers, C. Battaglia, J.-P. Becker, {\it et~al.}, {\em Pathways
  to electrochemical solar-hydrogen technologies}, Energy \& Environmental
  Science {\bf 11},  2768  (2018).

\bibitem{jakob2012}
J. Kibsgaard, Z. Chen, B. Reinecke~N., and T. Jaramillo~F., {\em Engineering
  the surface structure of {{MoS$_2$}} to preferentially expose active edge
  sites for electrocatalysis}, Nature Materials {\bf 11},  963   (2012).

\bibitem{lu2014ultrahigh}
Z. Lu, W. Zhu, X. Yu, H. Zhang, Y. Li, X. Sun, X. Wang, H. Wang, J. Wang, J.
  Luo, X. Lei, and L. Jiang, {\em Ultra-high hydrogen evolution performance of
  under-water superaerophobic {{MoS$_2$}} nanostructured electrodes}, Advanced
  Materials {\bf 26},  2683  (2014).

\bibitem{jiang2015}
Y. Li, H. Zhang, T. Xu, Z. Lu, X. Wu, P. Wan, X. Sun, and L. Jiang, {\em
  Under-water superaerophobic pine-shaped {{Pt}} nanoarray electrode for
  ultrahigh-performance hydrogen evolution}, Advanced Functional Materials {\bf
  25},  1737  (2015).

\bibitem{Taqieddin2017}
A. Taqieddin, R. Nazari, L. Rajic, and A. Alshawabkeh, {\em Physicochemical
  hydrodynamics of gas bubbles in two phase electrochemical systems}, Journal
  of The Electrochemical Society {\bf 164},  E448  (2017).

\bibitem{takanabe2017photocatalytic}
K. Takanabe, {\em Photocatalytic Water Splitting: Quantitative Approaches
  toward Photocatalyst by Design}, ACS Catalysis {\bf 7},  8006  (2017).

\bibitem{shen2008photocatalytic}
G. Shen, X.~H. Zhang, Y. Ming, L. Zhang, Y. Zhang, and J. Hu, {\em
  Photocatalytic induction of nanobubbles on {{TiO$_2$}} surfaces}, The Journal
  of Physical Chemistry C {\bf 112},  4029  (2008).

\bibitem{lakkaraju2013heat}
R. Lakkaraju, R.~J. Stevens, P. Oresta, R. Verzicco, D. Lohse, and A.
  Prosperetti, {\em Heat transport in bubbling turbulent convection},
  Proceedings of the National Academy of Sciences of the United States of
  America {\bf 110},  9237  (2013).

\bibitem{guzman2016}
D.~N. Guzman, Y. Xie, S. Chen, D.~F. Rivas, C. Sun, D. Lohse, and G. Ahlers,
  {\em {Heat-flux enhancement by vapour-bubble nucleation in Rayleigh-Benard
  turbulence}}, {Journal of Fluid Mechanics} {\bf {787}},  331  ({2016}).

\bibitem{guzman2016b}
D.~N. Guzman, T. Fraczek, C. Reetz, C. Sun, D. Lohse, and G. Ahlers, {\em
  {Vapour-bubble nucleation and dynamics in turbulent Rayleigh-Benard
  convection}}, {Journal of Fluid Mechanics} {\bf {795}},  60  ({2016}).

\bibitem{sur2018pool}
A. Sur, Y. Lu, C. Pascente, P. Ruchhoeft, and D. Liu, {\em Pool boiling heat
  transfer enhancement with electrowetting}, International Journal of Heat and
  Mass Transfer {\bf 120},  202  (2018).

\bibitem{wilhelmsen2015heat}
{\O}. Wilhelmsen, T.~T. Trinh, S. Kjelstrup, T.~S. van Erp, and D. Bedeaux,
  {\em Heat and mass transfer across interfaces in complex nanogeometries},
  Physical Review Letters {\bf 114},  065901  (2015).

\bibitem{cho2015turning}
H.~J. Cho, J.~P. Mizerak, and E.~N. Wang, {\em Turning bubbles on and off
  during boiling using charged surfactants}, Nature Communications {\bf 6},
  8599  (2015).

\bibitem{whitney1988mass}
G.~M. Whitney and C.~W. Tobias, {\em Mass-transfer effects of bubble streams
  rising near vertical electrodes}, AIChE Journal {\bf 34},  1981  (1988).

\bibitem{lohse2018}
D. Lohse, {\em Bubble puzzles: From fundamentals to applications}, Physical
  Review Fluids {\bf 3},  110504  (2018).

\bibitem{oguz1993dynamics}
H.~N. Oguz and A. Prosperetti, {\em Dynamics of bubble growth and detachment
  from a needle}, Journal of Fluid Mechanics {\bf 257},  111  (1993).

\bibitem{duhar2006dynamics}
G. Duhar and C. Colin, {\em Dynamics of bubble growth and detachment in a
  viscous shear flow}, Physics of Fluids {\bf 18},  077101  (2006).

\bibitem{kadyk2016}
T. Kadyk, D. Bruce, and M. Eikerling, {\em How to enhance gas removal from
  porous electrodes?}, Scientific Reports {\bf 6},  38780  (2016).

\bibitem{li2014coalescence}
D. Li, D. Jing, Y. Pan, W. Wang, and X. Zhao, {\em Coalescence and stability
  analysis of surface nanobubbles on the polystyrene/water interface}, Langmuir
  {\bf 30},  6079  (2014).

\bibitem{chan2015coalescence}
C.~U. Chan, M. Arora, and C.-D. Ohl, {\em Coalescence, growth, and stability of
  surface-attached nanobubbles}, Langmuir {\bf 31},  7041  (2015).

\bibitem{zhang2006removal}
X.~H. Zhang, G. Li, N. Maeda, and J. Hu, {\em Removal of induced nanobubbles
  from water/graphite interfaces by partial degassing}, Langmuir {\bf 22},
  9238  (2006).

\bibitem{shin2015growth}
D. Shin, J.~B. Park, Y.-J. Kim, S.~J. Kim, J.~H. Kang, B. Lee, S.-P. Cho, B.~H.
  Hong, and K.~S. Novoselov, {\em Growth dynamics and gas transport mechanism
  of nanobubbles in graphene liquid cells}, Nature communications {\bf 6},
  6068  (2015).

\bibitem{lv2017morphological}
P. Lv, Y. Xiang, Y. Xue, H. Lin, and H. Duan, {\em Morphological bubble
  evolution induced by air diffusion on submerged hydrophobic structures},
  Physics of Fluids {\bf 29},  032001  (2017).

\bibitem{xue2015morphology}
Y. Xue, P. Lv, Y. Liu, Y. Shi, H. Lin, and H. Duan, {\em Morphology of gas
  cavities on patterned hydrophobic surfaces under reduced pressure}, Physics
  of Fluids {\bf 27},  092003  (2015).

\bibitem{thoroddsen2005}
S. Thoroddsen, T. Etoh, K. Takehara, and N. Ootsuka, {\em {On the coalescence
  speed of bubbles}}, {Physics of Fluids} {\bf {17}},  071703  ({2005}).

\bibitem{zhang2008satellite}
F. Zhang and S. Thoroddsen, {\em Satellite generation during bubble
  coalescence}, Physics of Fluids {\bf 20},  022104  (2008).

\bibitem{soto2018coalescence}
{\'A}.~M. Soto, T. Maddalena, A. Fraters, D. van~der Meer, and D. Lohse, {\em
  Coalescence of diffusively growing gas bubbles}, Journal of Fluid Mechanics
  {\bf 846},  143  (2018).

\bibitem{boreyko2009self}
J.~B. Boreyko and C.-H. Chen, {\em Self-propelled dropwise condensate on
  superhydrophobic surfaces}, Physical Review Letters {\bf 103},  184501
  (2009).

\bibitem{zhang2009satellite}
F. Zhang, E. Li, and S.~T. Thoroddsen, {\em Satellite formation during
  coalescence of unequal size drops}, Physical Review Letters {\bf 102},
  104502  (2009).

\bibitem{wisdom2013self}
K.~M. Wisdom, J.~A. Watson, X. Qu, F. Liu, G.~S. Watson, and C.-H. Chen, {\em
  Self-cleaning of superhydrophobic surfaces by self-propelled jumping
  condensate}, Proceedings of the National Academy of Sciences of the United
  States of America {\bf 110},  7992  (2013).

\bibitem{liu2014numerical}
F. Liu, G. Ghigliotti, J.~J. Feng, and C.-H. Chen, {\em Numerical simulations
  of self-propelled jumping upon drop coalescence on non-wetting surfaces},
  Journal of Fluid Mechanics {\bf 752},  39  (2014).

\bibitem{enright2014coalescing}
R. Enright, N. Miljkovic, J. Sprittles, K. Nolan, R. Mitchell, and E.~N. Wang,
  {\em How coalescing droplets jump}, ACS Nano {\bf 8},  10352  (2014).

\bibitem{mouterde2017antifogging}
T. Mouterde, G. Lehoucq, S. Xavier, A. Checco, C.~T. Black, A. Rahman, T.
  Midavaine, C. Clanet, and D. Qu{\'e}r{\'e}, {\em Antifogging abilities of
  model nanotextures}, Nature Materials {\bf 16},  658  (2017).

\bibitem{mouterde2017merging}
T. Mouterde, T.-V. Nguyen, H. Takahashi, C. Clanet, I. Shimoyama, and D.
  Qu{\'e}r{\'e}, {\em How merging droplets jump off a superhydrophobic surface:
  Measurements and model}, Physical Review Fluids {\bf 2},  112001  (2017).

\bibitem{lecointre2019ballistics}
P. Lecointre, T. Mouterde, A. Checco, C.~T. Black, A. Rahman, C. Clanet, and D.
  Qu{\'e}r{\'e}, {\em Ballistics of self-jumping microdroplets}, Physical
  Review Fluids {\bf 4},  013601  (2019).

\bibitem{weon2012coalescence}
B.~M. Weon and J.~H. Je, {\em Coalescence preference depends on size
  inequality}, Physical Review Letters {\bf 108},  224501  (2012).

\bibitem{kim2015coalescence}
Y. Kim, S.~J. Lim, B. Gim, and B.~M. Weon, {\em Coalescence preference in
  densely packed microbubbles}, Scientific Reports {\bf 5},  7739  (2015).

\bibitem{weon2008coherent}
B. Weon, J. Je, Y. Hwu, and G. Margaritondo, {\em A coherent synchrotron
  \protect{X}-ray microradiology investigation of bubble and droplet
  coalescence}, Journal of Synchrotron Radiation {\bf 15},  660  (2008).

\bibitem{chen2017spatial}
R. Chen, H.~W. Yu, L. Zhu, R.~M. Patil, and T. Lee, {\em Spatial and temporal
  scaling of unequal microbubble coalescence}, AIChE Journal {\bf 63},  1441
  (2017).

\bibitem{le-the2018large-scale}
H. Le-The, E. Berenschot, R. Tiggelaar, N. Tas, A. van~den Berg, and J. Eijkel,
  {\em Large-scale fabrication of highly ordered sub-20 nm noble metal
  nanoparticles on silica substrates without metallic adhesion layers},
  Microsystems \& Nanoengineering {\bf 4},  4  (2018).

\bibitem{lv2017growth}
P. Lv, H. Le~The, J. Eijkel, A. Van~den Berg, X. Zhang, and D. Lohse, {\em
  Growth and detachment of oxygen bubbles induced by gold-catalyzed
  decomposition of hydrogen peroxide}, The Journal of Physical Chemistry C {\bf
  121},  20769  (2017).

\bibitem{chen2009growth}
S.-L. Chen, C.-T. Lin, C. Pan, C.-C. Chieng, and F.-G. Tseng, {\em Growth and
  detachment of chemical reaction-generated micro-bubbles on micro-textured
  catalyst}, Microfluidics and Nanofluidics {\bf 7},  807  (2009).

\bibitem{magnaudet2000}
J. Magnaudet and I. Eames, {\em The motion of high-Reynolds-number bubbles in
  inhomogeneous flows}, Annual Review of Fluid Mechanics {\bf 32},  659
  (2000).

\bibitem{lamb1945Hydrodynamics}
S.~H. Lamb, {\em Hydrodynamics}, sixth edition ed. (Dover Publications, New
  York, 1945).

\bibitem{marmottant2006high}
P. Marmottant, M. Versluis, N. De~Jong, S. Hilgenfeldt, and D. Lohse, {\em
  High-speed imaging of an ultrasound-driven bubble in contact with a wall:
  "Narcissus" effect and resolved acoustic streaming}, Experiments in Fluids
  {\bf 41},  147  (2006).

\bibitem{versluis2010microbubble}
M. Versluis, D.~E. Goertz, P. Palanchon, I.~L. Heitman, S.~M. van~der Meer, B.
  Dollet, N. de~Jong, and D. Lohse, {\em Microbubble shape oscillations excited
  through ultrasonic parametric driving}, Physical Review E {\bf 82},  026321
  (2010).

\bibitem{phibbs1951hydrogen}
M. Phibbs and P.~A. Gigu{\`e}re, {\em Hydrogen peroxide and its analogues: I.
  density, refractive index, viscosity, and surface tension of deuterium
  peroxide--deuterium oxide solutions}, Canadian Journal of Chemistry {\bf 29},
   173  (1951).

\bibitem{Suppl}
 Supplemental Material for extended discussions on the effect of liquid
  viscosity on the jumping behaviors, and physical properties of solution and
  substrate  .

\bibitem{plesset1977bubble}
M.~S. Plesset and A. Prosperetti, {\em Bubble dynamics and cavitation}, Annual
  Review of Fluid Mechanics {\bf 9},  145  (1977).

\bibitem{stover1997bubble}
R.~L. Stover, C.~W. Tobias, and M.~M. Denn, {\em Bubble coalescence dynamics},
  AIChE Journal {\bf 43},  2385  (1997).

\bibitem{lohse2020}
D. Lohse and X. Zhang, {\em Physicochemical hydrodynamics of droplets out of
  equilibrium}, Nature Reviews Physics {\bf 2},  426  (2020).

\end{thebibliography}

\end{document}